\documentclass[a4paper,10pt,twoside]{cpc-hepnp}

\usepackage{multicol}
\usepackage{graphicx}
\usepackage{booktabs}
\usepackage{amssymb,bm,mathrsfs,bbm,amscd}
\usepackage[tbtags]{amsmath}
\usepackage{lastpage}
\usepackage{natbib}
\usepackage{multirow}

\begin{document}

\fancyhead[c]{\small Submitted to 'Chinese Physics C'}
\fancyfoot[C]{\small 0000000 -\thepage}

\footnotetext[0]{Received date Month Year}

\title{Measurement and Analysis of Fission Rates in a Spherical Mockup of Uranium and Polyethylene\thanks{Supported by Chinese Special Project for ITER (2010GB111002) }}

\author{%
      ZHU Tong-Hua$^{1,2;1)}$\email{zhutonghua@163.com}%
\quad YANG Chao-Wen$^{1}$
\quad LU Xin-Xin$^{2}$
\quad LIU Rong$^{2}$\\
\quad HAN Zi-Jie$^{2}$
\quad JIANG Li$^{2}$
\quad WANG Mei$^{2}$
}
\maketitle

\address{%
$^1$ Institute of Nuclear Science and Technology,  Sichuan University,  Chengdu 610065, China\\
$^2$ Institute of Nuclear Physics and Chemistry, China Academy of Engineering Physics, Mianyang 621900, China\\
}

\begin{abstract}
Measurements of the reaction rate distribution were carried out using two kinds of Plate Micro Fission Chamber(PMFC). The first is a depleted uranium chamber and the second an enriched uranium chamber. The material in the depleted uranium chamber is strictly the same as the material in the uranium assembly. With the equation solution to conduct the isotope contribution correction, the fission rate of $^{238}$U and $^{235}$U were obtained from the fission rate of depleted uranium and enriched uranium. And then, the fission count of $^{238}$U and $^{235}$U in an individual uranium shell was obtained. In this work, MCNP5 and continuous energy cross sections ENDF/BV.0 were used for the analysis of fission rate distribution and fission count. The calculated results were compared with the experimental ones. The calculation of fission rate of DU and EU were found to agree with the measured ones within 10\% except at the positions in polyethylene region and the two positions near the outer surface. Beacause the fission chamber was not considered in the calculation of the fission counts of $^{238}$U and $^{235}$U, the calculated results did not agree well with the experimental ones.
\end{abstract}



\footnotetext[0]{\hspace*{-3mm}\raisebox{0.3ex}{$\scriptstyle\copyright$}2013
Chinese Physical Society and the Institute of High Energy Physics
of the Chinese Academy of Sciences and the Institute
of Modern Physics of the Chinese Academy of Sciences and IOP Publishing Ltd}%

\begin{multicols}{2}

\section{Introduction}
In the new blanket concept of fusion-fission hybrid, depleted uranium was considered to be candidate for neutron breeding and energy generation, and water for heat transmission. The volume ratio of about 2.0 for uranium to hydrogen was selected. It is considered that such structure is the best one for energy production and fissile breeding, and also for heat transmission\cite{lab1,lab2,lab3}. For the feasibility of energy breeding and neutron multiplication in this blanket concept, the neutron transport and fission generation should be accurately predicted\cite{lab4}. To do this, the method, code and data are of fundamental importance in the concept research as in the design of fission plants, fusion devices\cite{lab5}. Therefore, a series of fission rate experiments on the assemblies having similar configuration with the hybrid blanket were planned. Now, a spherical assembly of depleted uranium and polyethylene in which similar ratio of uranium to hydrogen was selected had been setup to simulate the structure in blanket concept design and the fission rate experiment on the assembly by fission chamber had been conducted with D-T neutron generated at the center. For comparison with the experimental data, as in hybrid concept study, radiation transport tool MCNP5\cite{lab6} and nuclear data library ENDF/BV.0\cite{lab7} was adopted in this work to be checked.

\section{Experiment and procedure}
A neutronics experiment of fission rate distribution in an assembly of depleted uranium and polyethylene has been conducted with D-T neutron at the center. Fig.\ref{fig1} shows the arrangement between the experimental assembly and D-T neutron source. The experimental assembly was consist of two layers of polyethylene and three layers of depleted uranium. The dimensions of the assembly (outer radius/inner radius) is 30 cm(DU)25.4 cm(PE)23.34 cm(DU)19.4 cm(PE)18.1 cm(DU)13.1 cm. The spherical shells of polyethylene were newly constructed while the depleted uranium shells were constructed in the large depleted uranium experiments\cite{lab8}. There are six channels in the assembly, one of which is for detector and one for drift tube while the others was just filled with cylindrical depleted uranium blocks and depleted uranium sleeves. The uranium sleeves were also used in the channel for detector to keep the blocks of uranium and polyethylene from dropping down and to install the detector at a specific position. The material in the sleeves are keeped strictly the same to the material in the shells by combining various blocks with different thickness of 0.2 cm to 5.0 cm except for the detector.

\begin{center}
\includegraphics[width=8cm]{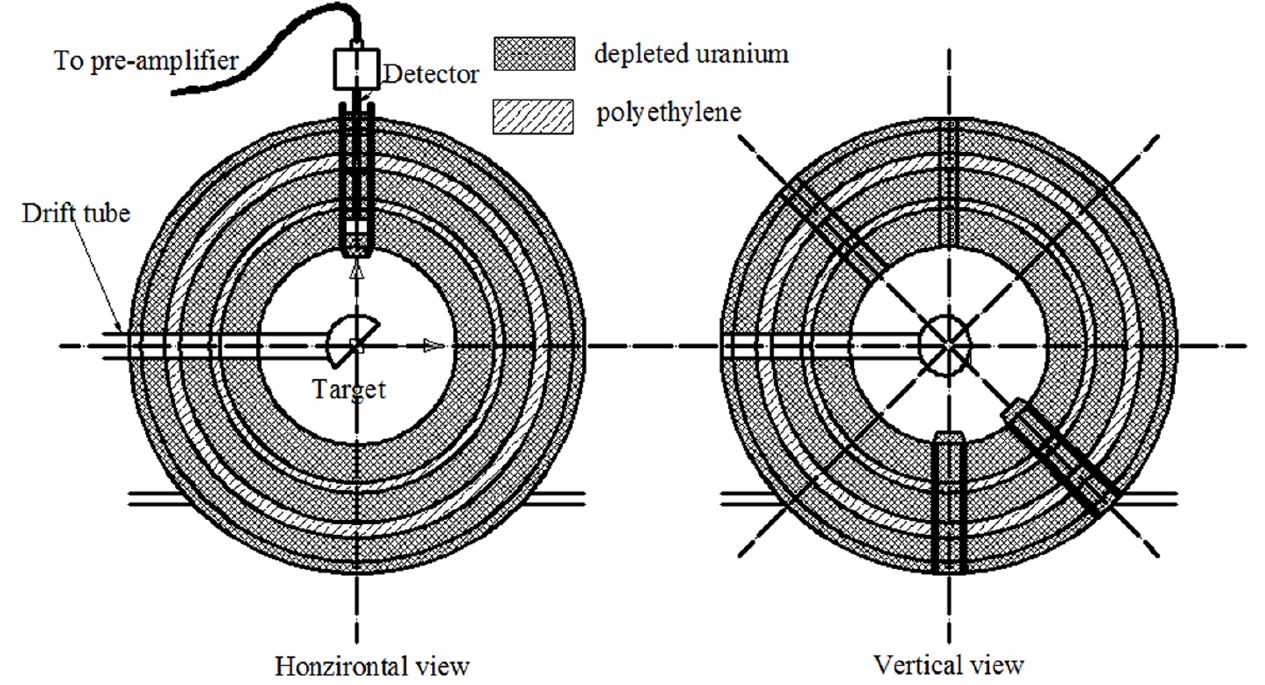}
\figcaption{\label{fig1}   The schematical view of spherical assembly of depleted uranium and polyethylene. }
\end{center}

D-T neutrons were generated by bombardment of 250 keV and 100 uA deuteron onto a tritiated titanium target (90 GBq). The intensity was about 1.5¡Á10$^{10}$ neutrons/s. The fluence of D-T neutrons were monitored with counting the associated 3.5 MeV alpha particles by a silicon surface barrier detector at the angle of 178.2¡ãto the incident deuteron beam and the experimental error of the fluence of D-T neutrons was about 2.5\%\cite{lab9}. Depleted uranium and Enriched uranium Plate-Micro-Fission-Chambers(PMFC) were used to measure the fission rate along the channel as a function of distance from the core of the assembly to the detector position. The atomic densities of fissile material in the PMFCs have been accurately measured from the alpha particle emitted from the fissile material before the PMFCs were fabricated. Table~\ref{tab1} shows the characterizations of PMFCs, including atomic densities, dimensions, active area and so on. The sources of the uncertainties in measured results were summaried in Table~\ref{tab2}.

\end{multicols}
\ruleup
\begin{center}
\tabcaption{ \label{tab1}  The characterization of the PMFCs.}
\footnotesize
\begin{tabular*}{170mm}{c@{\extracolsep{\fill}}ccccccc}
\toprule PMFCs&Nuclide&Number/atoms&\%Uncertainty&Detector size/cm& Active area/$cm^{2}$&\%Efficiency\\
\hline
DU&$^{238}$U & 3.1$\times{10}$$^{18}$ &1.3&$\phi$ 3$\times{2}$&4.52&95.64\\
\hphantom{00} & $^{235}$U & 1.3$\times{10}$$^{16}$ & \hphantom{00} & \hphantom{00}  & \hphantom{00}  & \hphantom{00}\\
EU&$^{238}$U& 2.1$\times{10}$$^{17}$&2.4&$\phi$ 3$\times{2}$ & 4.52&96.63\\
\hphantom{00} & $^{235}$U& 2.3$\times{10}$$^{18}$ &\hphantom{00} & \hphantom{00} & \hphantom{00}  & \hphantom{00}\\
\bottomrule
\end{tabular*}
\end{center}
\vspace{5mm}
\ruledown

\begin{multicols}{2}

\begin{center}
\tabcaption{ \label{tab2}  The sources of the uncertainties in fission rate measurement.}
\footnotesize
\begin{tabular*}{80mm}{l@{\extracolsep{\fill}}cc}
\toprule Sources&\% \\
\hline
Absolute neutron yield & 2.5\\
\hphantom{00}Statistics of alpha counts & 0.5\\
\hphantom{00}Solid angle viewed by the detector & 1.7\\
\hphantom{00}Correction for neutron anistropy & 1.2\\
\hline
Fission chamber measurement\ & 1.64(DU)/2.6(EU)\\
\hphantom{00}Atomic number density & 1.3(DU)/2.4(EU)\\
\hphantom{00}Counting statistics & 1.0\\
\hline
Total uncertainty & 2.99(DU)/3.61(EU)\\
\bottomrule
\end{tabular*}
\vspace{0mm}
\end{center}
\vspace{0mm}

Measured fission rates have been estimated by the following ~Eq.~(\ref{eq1})
\begin{eqnarray}
\label{eq1}
f(r)=\frac{N_f}{{A}{N}\eta}
\end{eqnarray}

where $f(r)$ is the fission rate of depleted uranium(DU) or enriched uranium(EU) at a specific position. r is the radius and $N_{f}$, $A$,$N$,$\eta$ are the sum of yields obtained by PMFCs in which the fission reactions occured with $^{238}$U,$^{235}$U,$^{234}$U and so on, the atom number of all uranium isotopes in the PMFCs, the total amount of neutrons emitted from D-T reaction and the efficiency of fission fragments shaping a pulse in PMFCs, respectively. In addition, $N_{f}$ has been corrected by dead time.


For the $^{234}$U contribution as small as 1.0\% in DU PFMC and EU PFMC, there is just $^{238}$U and $^{235}$U which have to be considered for contribution to the measured fission yields in ~Eq.~(\ref{eq1}). So in a specific DU or EU PFMC, the fission yields $N_{fd}$ or $N_{fe}$ are formed mainly by that of $^{238}$U $N_{f8d}$ or $N_{f8e}$ and that of $^{235}$U $N_{f5d}$ or  $N_{f5e}$. Then based on the ~Eq.~(\ref{eq1}), the fission yields $N_{fd}$ or $N_{fe}$ can be given as following equations.
\begin{eqnarray}
\label{eq4}
\begin{aligned}
N_{fd}&=N_{f8d}+N_{f5d}\\
    &=f_{8}{A_{8d}}{N}\eta+f_{5d}{A_{5d}}{N}\eta\ \\
    &=(f_{8}{A_{8d}}+f_{5}{A_{5d}}){N}\eta\ \\
    &=(f_{8}{A_d}K_{8d}+f_{5}{A_d}K_{5d}){N}\eta\ \\
    &=(f_{8}K_{8d}+f_{5}K_{5d}){A_d}{N}\eta\
\end{aligned}
\end{eqnarray}
\begin{eqnarray}
\label{eq5}
\begin{aligned}
N_{fe}&=N_{f8e}+N_{f5e}\\
    &=f_{8}{A_{8e}}{N}\eta+f_{5e}{A_{5e}}{N}\eta\ \\
    &=(f_{8}{A_{8e}}+f_{5}{A_{5e}}){N}\eta\ \\
    &=(f_{8}{A_e}{K_{8e}}+f_{5}{A_e}{K_{5e}}){N}\eta\ \\
    &=(f_{8}{K_{8e}}+f_{5}{K_{5e}}){A_e}{N}\eta\
\end{aligned}
\end{eqnarray}

Here, $f_{8}$ is the fission rate of $^{238}$U fission reaction, $f_{5}$ is the sum of fission rate of $^{235}$U fission reaction, $f_{d}$ is the fission rate of depleted uranium, $f_{e}$ is the fission rate of enriched uranium, $K_{8d}$ and $K_{5d}$ is the atom percents of $^{238}$U and $^{235}$U in depleted uranium while $K_{8e}$ and $K_{5e}$ is the atom percents of $^{238}$U and $^{235}$U in enriched uranium. $A_{d}$ is the number of fission nuclide in DU chamber while $A_{e}$ is that in EU chamber. In this work, $K_{8d}$ and $K_{5d}$  were 0.9958 and 0.0042 while $K_{8e}$ and $K_{5e}$ were 0.0863 and 0.9137.

According to equation (1,2,3), the relations of the fission rate of DU or EU and the fission rate of $^{238}$U and $^{235}$U in DU or EU can be presented by following equations.
\begin{eqnarray}
\label{eq6}
\begin{aligned}
f_{d}&=\frac{N_{fd}}{{A_d}{N}\eta}\\
     &=f_{8}\times{K_{8d}}+f_{5}\times{K_{5d}}.
\end{aligned}
\end{eqnarray}
\begin{eqnarray}
\label{eq7}
\begin{aligned}
f_{e}&=\frac{N_{fe}}{{A_e}{N}\eta}\\
&=f_{8}\times{K_{8e}}+f_{5}\times{K_{5e}}.
\end{aligned}
\end{eqnarray}

As we can see, there are two equation with two unknown parameters. Based on fission rate of DU $f_{d}$ and EU $f_{e}$, the fission rate of $^{238}$U and $^{235}$U can be obtained, the method was named equation method.

All the fission rate results are normalized to one atom, one source neutron.Then based on the fission rate of $^{238}$U and $^{235}$U, the fission counts of $^{238}$U and $^{235}$U in a specific uranium layer of the assembly can be obtained by integration method. The following formula shows the integration equation to get the fission count of $^{238}$U,

\begin{eqnarray}
\label{eq8}
P_{f8}=(2\pi\rho\times{\frac{K_{8d}}{238})}\times{\int\nolimits\int\nolimits{r^{2}f_{8}(r)sin{\theta}{\rm d}{\theta}{\rm d}{r}}}
\end{eqnarray}
\begin{eqnarray}
\label{eq9}
P_{f8}=(4\pi\rho\times{\frac{K_{8d}}{238})}\times{\int\nolimits\int\nolimits{r^{2}f_{8}(r){\rm d}{r}}}.
\end{eqnarray}

where $\rho$ is the density of the depleted uranium assemblies; r is the distance of the measuring position to the core, ranging from 0 to R, R is the outer radius of the assemblies; $\theta$ is the angle of the measuring position to incident D$^{+}$ beam and $K_{8d}$ has the same meaning in ~Eq.~(\ref{eq4}), the fission counts of $^{235}$U can be given in the simillar equation. According to the experimental analysis  before, the fission rate along the channel which is perpendicular to the incident D$^{+}$ beam is almost the average of all fission rate which at the assymetric angle to the experimental channel. So the ~Eq.~(\ref{eq8}) was simplified to be ~Eq.~(\ref{eq9}). Then with ORIGIN 6.0 code and trapezoidal area method, the fission count was obtained. The uncertainty of fission count is about 3.2\% for $^{238}$U and 4.2\% for $^{235}$U. The contribution to the quoted uncertainty coming from the fission rate of $^{238}$U (3.0\%) and $^{235}$U (4.0\%), the distance of the measuring position to the core (1.0\%).
\section{Monte carlo analysis}
Analyses of the fission rate experiment were carried out with the Monte-Carlo code MCNP 5\cite{lab6} and attached continuous energy cross section ENDF/BV.0\cite{lab7}. The experimental configuration was modeled in detail, including target chamber, drift tube, void in sleeves, the sheet for detector and so on . The surface crossing estimator was used. Neutron histories were accumulated to obtain a good statistical accuracy, less than 2.0\%. In the calculation, an isotropic distribution of source neutrons was assumed. By filling all the channels with uranium block in the calculation model and a point neutron source with 14.1 MeV was placed in the center, fission counts in individual uranium shell were obtained, and the influence of materials in detectors was also analyzed.
\section{Results and discussion}
\subsection{Fission rate distribution}
Fig.~\ref{fig2} shows experimental and calculated results of depleted uranium and enriched uranium fission rate as a function of distance from the core to measuring position. In the figures, DU means the material in the shell is depleted uranium and PE means the material in the shell is polyethylene. The experimental error band was smaller than the size of the dots showing the experimental results, and the calculation ones also. The depleted uranium fission rate in which the $^{238}$U contribution is as high as 90\% decrease with the distance from the core increasing, and the enhancement of fission rate around the polyethylene region was also indicated from the two platform on the curve. As for enriched uranium fission rate in which the $^{235}$U contribution is the main content seems very sensible to the distance and also to the polyethylene, this is caused by the change of slow neutron strength in the assembly which is sensible to the average free path and the elastic scattering of Carbon and Hydrogen.
\begin{center}
\includegraphics[width=6cm]{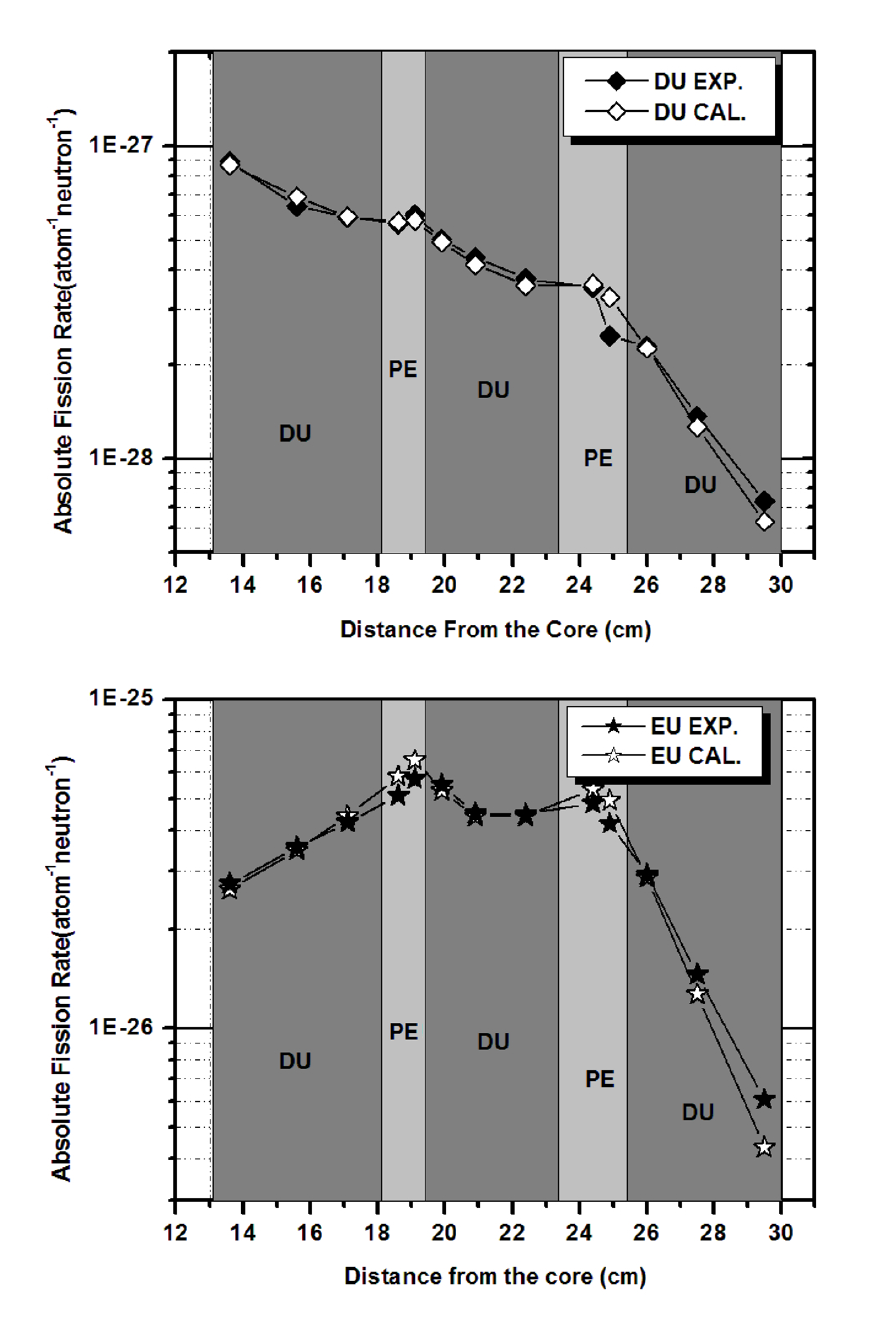}
\figcaption{\label{fig2}   The measured and calculated fission rate of DU and EU}
\end{center}

The tendency of fission rate distribution as a function of the distance from the core is well repeated by the calculation, and the Calculated results to Experimental ones (C/Es) are limited to the range of ¡À1.1 while the results in polyethylene and at the positions near the outer surface is beyond this range (see Fig.3). The main reason of the overestimation around the polyethylene region are considered as the contribution of neutron resonance while  underestimation at the further point on the assembly was due to contribution of room returned neutrons.

\begin{center}
\includegraphics[width=5.5cm]{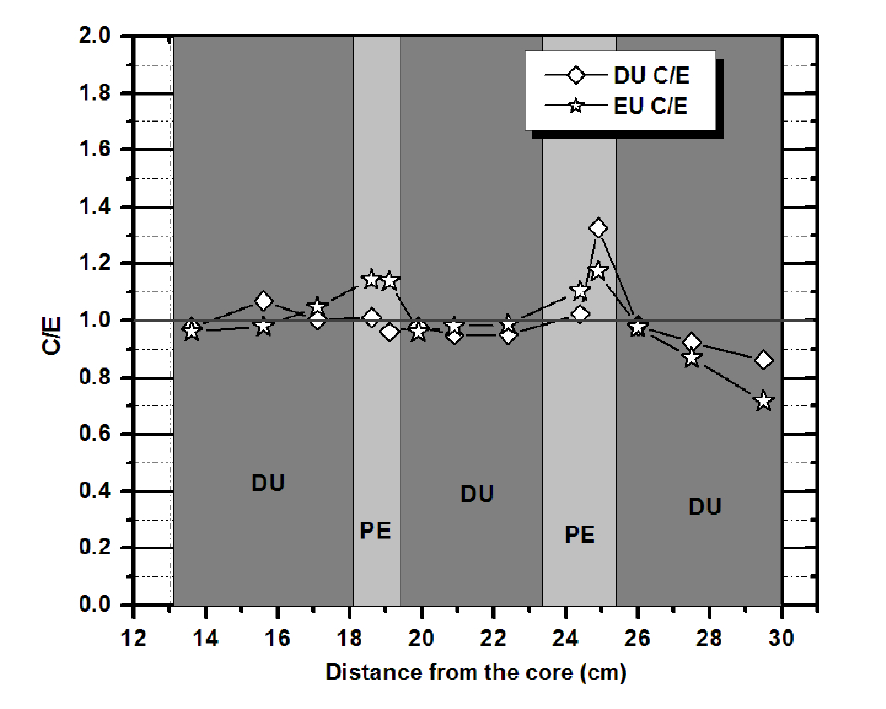}
\figcaption{\label{fig3}   C/Es of the fission rate of DU and EU in the assembly}
\end{center}

\subsection{Fission counts}
The ideal spherical shell model was used for calculation of $^{238}$U or $^{235}$U fission counts in each uranium shell, in which the detector channel was filled with uranium block while the others parts is the same as the model in  calculation of $^{238}$U or $^{235}$U fission rate distribution. In this paper, the fission count were calculated by using two methods: a) calculation of the U fission rate distribution at the same positions as the ones in experiment by using surface cross estimator and then the fission count was got with the integration method as in experiment; b) directly by using the three individual uranium shell as volume estimator.

\begin{center}
\includegraphics[width=7cm]{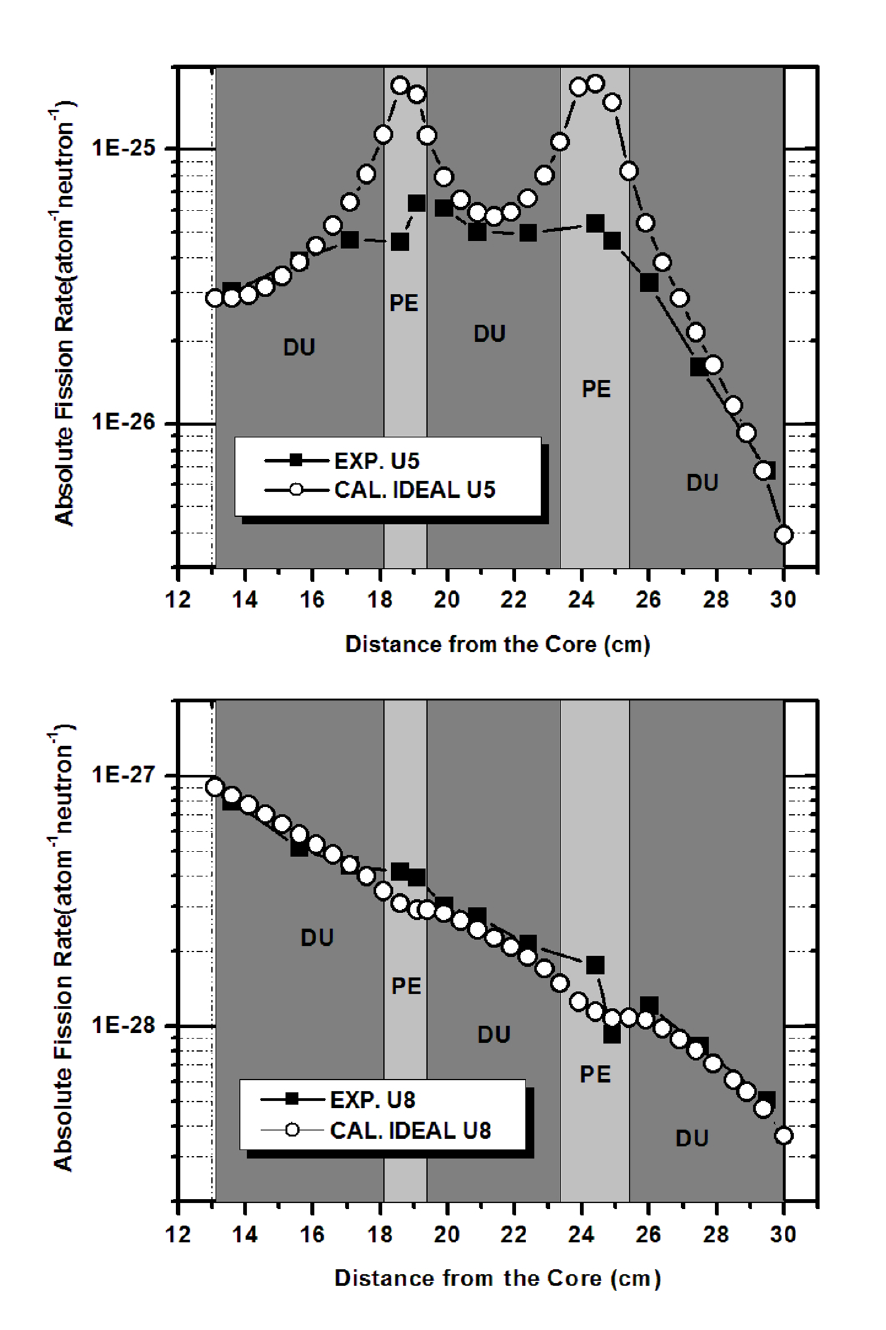}
\figcaption{\label{fig4}   The comparison of the calculated results of $^{238}$U and $^{235}$U fission rate with method a) and the experimental ones}
\end{center}

Fig.~\ref{fig4} shows the comparison of the calculated results of $^{238}$U or $^{235}$U fission rate from method a) and that deduced from the experimental DU and EU fission rate distribution by equation (\ref{eq6},\ref{eq7}). The calculation of $^{238}$U fission rate agree well with the experiment in the uranium shell while large difference exist in the $^{235}$U results in the uranium region. The similar comparison results between calculation and experiment fission count indicated in Fig.~\ref{fig5}, where the calculated fission counts results with method a) and method b) and the ones from experimental measurement are presented. The discrepancy between the calculated fission count results from method a) and method b) was within 10\% while large discrepancy exist in the comparison of calculated results and experimental results. The difference between calculation model and experimental setup may be the main cause of the discrepancy between calculation and experimental results, especially the fission chamber was not considered in the calculation. This caused the neutron spectrum change from experiment to calculation especially in the PE region. For $^{238}$U with threshold of about 1.0 MeV, the results obtained by fission chamber were almost the same as the ones with point estimator in the ideal spherical shell. On the other hand, for $^{235}$U which is sensitive to slow neutrons, the large discrepancy between the experimental results and the calculation ones indicated around the polyethylene region.

\subsection{The influence of materials in detectors}
In the equation solution to get the $^{238}$U fission rate from the depleted uranium and the $^{235}$U fission rate from the enriched uranium results, it is assumed that the neutron field and the reaction probability of $^{238}$U in DU and in EU is the same or almost the same. For getting the $^{235}$U fission rate, the same condition was assumed. Because of the large cross sections of $^{235}$U and slow neutrons, the neutron energy spectra at the same position may be different between depleted uranium condition and enriched uranium condition. To evaluate the influence of material in detectors on the fission rate results, especially when the enriched uranium detector was used to get the fission rate of $^{235}$U. Two different models were established, in which depleted uranium fission chamber and enriched uranium fission chamber were included at the measuring position respectively. The fission rate of $^{238}$U, $^{235}$U were calculated in these two conditions, and the results were compared in Table~\ref{tab3}. No significant difference about 3\% was found.

\begin{center}
\includegraphics[width=6cm]{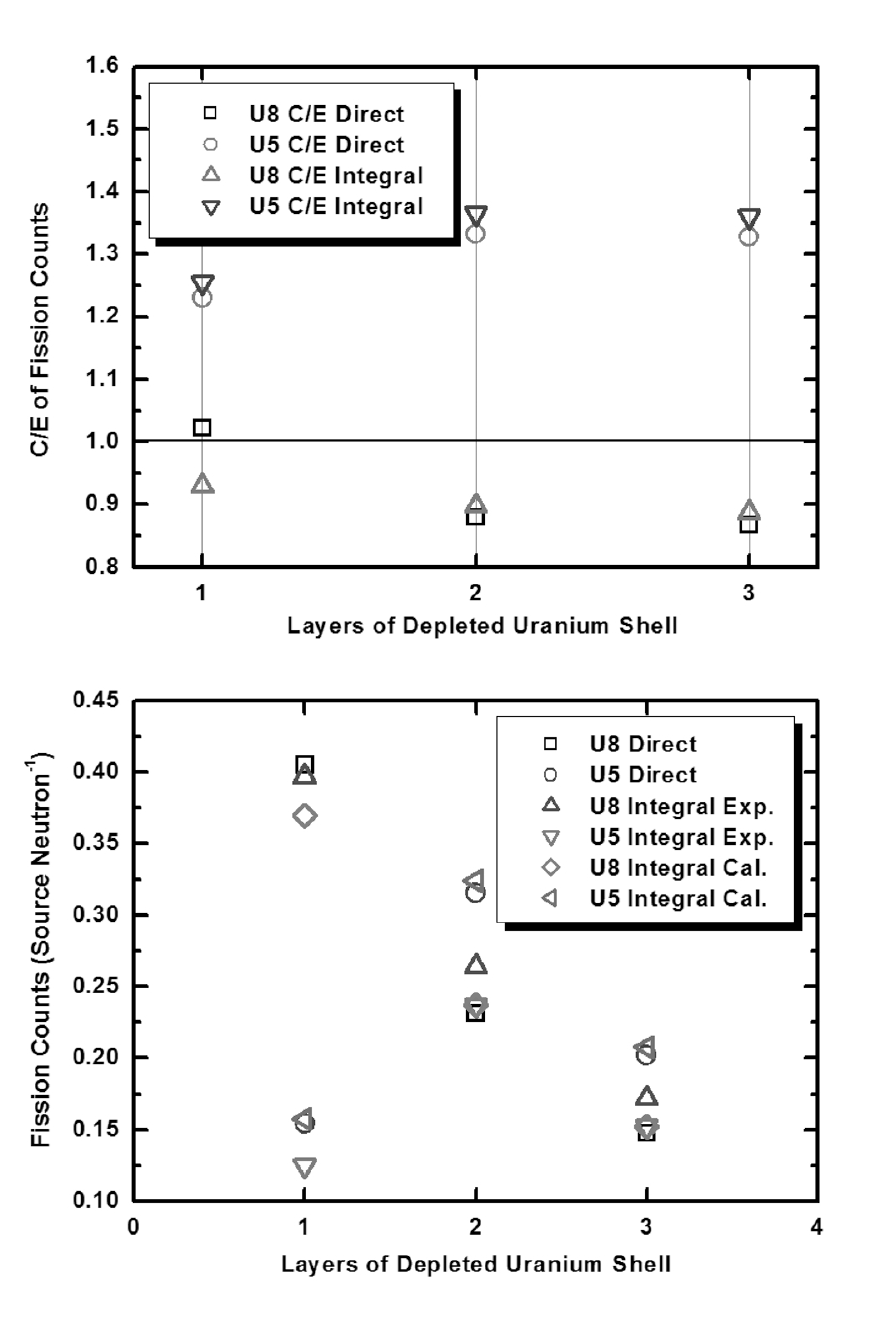}
\figcaption{\label{fig5}   The fission counts of $^{238}$U and $^{235}$U in individual uranium shell}
\end{center}

\end{multicols}
\ruleup
\begin{center}
\tabcaption{ \label{tab3}  The fission rate of $^{238}$U and $^{235}$U in two detector condition (atom$^{-1}$neutron$^{-1}$.)}
\footnotesize
\begin{tabular*}{170mm}{l@{\extracolsep{\fill}}ccccc}
\toprule distance from the core& $f_{5}$ in EU PMFC & $f_{5}$ in DU PMFC & $f_{8}$ in EU PMFC & $f_{8}$ in DU PMFC\\
\hline
13.6 cm & 2.80$\times{10}$$^{-26}$ &2.85$\times{10}$$^{-26}$& 7.65$\times{10}$$^{-28}$& 7.53$\times{10}$$^{-28}$\\
18.6 cm & 6.42$\times{10}$$^{-26}$ &6.30$\times{10}$$^{-26}$& 2.92$\times{10}$$^{-28}$& 2.95$\times{10}$$^{-28}$\\
24.9 cm & 5.35$\times{10}$$^{-26}$ &5.20$\times{10}$$^{-26}$& 1.04$\times{10}$$^{-28}$& 1.03$\times{10}$$^{-28}$\\
13.6 cm & 4.67$\times{10}$$^{-27}$ &4.78$\times{10}$$^{-27}$& 4.20$\times{10}$$^{-29}$& 4.25$\times{10}$$^{-29}$\\
\bottomrule
\end{tabular*}
\end{center}
\vspace{5mm}
\ruledown

\begin{multicols}{2}

\section{Conclusion}
The fission rate measurement have been conducted in the spherical assembly of depleted uranium and polyethylene, the experiment was introduced in detail and the results are presented with the calculated ones. The following facts are found:
1)	The calculation well repeated the distribution of DU and EU fission rate in the experimental assembly. Except the fission rates at the two positions near the outer surface of the assembly and the results at the positions in the polyethylene area, all the calculations of fission rate for DU and EU are agreed with the experimental ones within 10\%. The main reason of the overestimation around the polyethylene region are considered as the contribution of neutron resonance while  underestimation at the further point on the assembly was due to contribution of room returned neutrons.
2)	Because of the difference between calculation model and experimental setup, the discrepancy of fission counts for $^{238}$U with threshold of about 1.0 MeV is about 10\% while the overestimation of fission counts for $^{235}$U which is sensitive to slow neutrons in the individual uranium shell is up to 35\% around the polyethylene region. On the other hand, the difference between the calculated results of $^{238}$U and $^{235}$U fission counts from method a) and method b) was within 10\%.To get more accurate experimental results of fission counts in each shells, more measuring points are needed, especially around the boundary of uranium and polyethylene.
3)	More experiment are needed for the fission blanket concept design. The MCNP5 code and ENDF/B-V.0 data can be used in the neutronics design for hybrid concept with neutron energy above 1.0 MeV. More validations would be considered to check the calculations.\\

\acknowledgments{Authors would like to express their sincere thanks to Professor Wang Dalun for their helpful discussion and strong support in this work and to Lou Benchao, Zhang Qinlong for good operation of neutron generator.}
\end{multicols}

\vspace{-1mm}
\centerline{\rule{80mm}{0.1pt}}
\vspace{2mm}

\begin{multicols}{2}

\end{multicols}

\clearpage

\end{document}